\RequirePackage{ifpdf}

\documentclass{PoS}

 \usepackage{graphicx}
 \usepackage{epstopdf}

\title{All-sky signals from recombination to reionization with the SKA}

\ShortTitle{All-sky cosmological signals}

\author{Ravi Subrahmanyan\\
	Raman Research Institute, C V Raman Avenue, Sadashivanagar, Bangalore 560080, India\\
        E-mail: \email{rsubrahm@rri.res.in} }

\author{N Udaya Shankar\\
        Raman Research Institute, C V Raman Avenue, Sadashivanagar, Bangalore 560080, India\\
       E-mail: \email{uday@rri.res.in}}

\author{Jonathan Pritchard\\
        Imperial College London, Astrophysics, Blackett Laboratory, Prince Consort Road, London SW7 2AZ, UK\\
        E-mail: \email{j.pritchard@imperial.ac.uk}}

\author{\speaker{Harish K Vedantham}\\
       Kapteyn Astronomical Institute, Landleven 12, 9747 AD Groningen, The Netherlands  \\
        E-mail: \email{h.k.vedantham@rug.nl}}

\abstract{
Cosmic evolution in the hydrogen content of the Universe through recombination and up to the end of reionization is expected to be revealed as subtle spectral features in the uniform extragalactic cosmic radio background.  The redshift evolution in the excitation temperature of the 21-cm spin flip transition of neutral hydrogen appears as redshifted emission and absorption against the cosmic microwave background.  The precise signature of the spectral trace from cosmic dawn and the epoch of reionization are dependent on the spectral radiance, abundance and distribution of the first bound systems of stars and early galaxies, which govern the evolution in the spin-flip level populations.   Redshifted 21~cm from these epochs when the spin temperature deviates from the temperature of the ambient relic cosmic microwave background results in an all-sky spectral structure in the 40--200~MHz range, almost wholly within the band of SKA-Low.  Another spectral structure from gas evolution is redshifted recombination lines from epoch of recombination of hydrogen and helium; the weak all-sky spectral structure arising from this event is best detected at the upper end of the 350--3050~MHz band of SKA-mid.  Total power spectra of SKA interferometer elements form the measurement set for these faint signals from recombination and reionization; the inter-element interferometer visibilities form a calibration set.  The challenge is in precision polarimetric calibration of the element spectral response and solving for additives and unwanted confusing leakages of sky angular structure modes into spectral modes.  Herein we discuss observing methods and design requirements that make possible these all-sky SKA measurements of the cosmic evolution of hydrogen.   
}

\FullConference{
Advancing Astrophysics with the Square Kilometre Array\\
June 8-13, 2014\\
Giardini Naxos, Italy}

\newcommand{\skipthis}[1]{}

\newcommand{\lya}{Ly$\alpha$ }

\begin{document}

\section{Introduction: the science case}


From the time of big bang nucleosynthesis hydrogen gas makes up $\sim$75\% of the mass of the baryons in the Universe. This makes atomic lines of neutral hydrogen in the intergalactic medium an ideal tracer of the evolution and history of the Universe over cosmic time. In this section, we review the physics affecting hydrogen over cosmic time and the resulting signals that SKA can measure. A number of useful reviews on this topic exist, including \cite{2006PhR...433..181F,2012RPPh...75h6901P,Sunyaev:2009fk}.

Until the Universe cools to a temperature of $\sim0.3\ {\rm eV}$ photons from the radiation background keep hydrogen ionised. At a redshift of $z\approx1100$ hydrogen recombines leading to physical decoupling of the photons.  From that point on, the majority of photons free stream without scattering to the present day when they appear as the relic cosmic microwave background (CMB). Fully ionised helium (He~{\sc iii}), with its higher binding energy, recombines somewhat earlier first to He~{\sc ii} (at $z \approx $ 5000-8000) and then to He~{\sc i} (at $z \approx $ 1600-3500).

Although the photons decouple from the baryons at $z\approx1100$, the large photon-to-baryon ratio (approximately $10^{10}$ photons per baryon) means that scattering of photons from the small residual ionised fraction of free electrons ($x_e\sim10^{-5}$) maintains thermal equilibrium between the free electrons and the CMB. Collisions between free electrons and the remainder of the baryons maintains $T_K=T_{\rm CMB}$. This thermal coupling eventually ends at a redshift $z\approx200$ as dilution of the gas with Hubble expansion reduces the photon-electron scattering rate below a critical threshold. With thermal decoupling, the gas begins to cool adiabatically with expansion and its temperature scales as $T_K\propto(1+z)^2$, cooling faster than the CMB, which evolves as $T_{\rm CMB}\propto(1+z)$. For the first time in the Universe's evolution there exist two separate temperature scales that dictate the level populations in the atomic gas - the gas kinetic temperature $T_K$ and the CMB temperature $T_{\rm CMB}$. This temperature difference creates the opportunity for an interesting 21~cm signal.

Key to understanding the 21~cm signal is the evolution of the spin temperature $T_S$, which describes the ratio of atoms in the upper and lower hyperfine level $n_1/n_0=3\exp(-T_\star/T_S)$, where $T_\star\equiv hc/k\lambda_{21 \rm{cm}}=0.068\,\rm{K}$. The spin temperature is set by the relative rate of spin-flip transitions driven by interaction with CMB photons, scattering \lya photons, or collisions between the mix of neutral hydrogen atoms, free protons and free electrons. Formally it is determined by
\begin{equation}
T_S^{-1}=\frac{T_{\rm CMB}^{-1}+x_\alpha T_\alpha^{-1}+x_c T_K^{-1}}{1+x_\alpha+x_c},
\end{equation}
where $T_\alpha$ is the color temperature of the \lya radiation field at
the \lya frequency and $x_c$, $x_\alpha$ are coupling coefficients due to atomic collisions and scattering of Ly$\alpha$ photons, respectively.  $T_\alpha$ is closely coupled to $T_{K}$ by recoil during repeated scattering and the spin temperature becomes strongly coupled to the gas temperature when $x_{\rm{tot}}\equiv x_c+x_\alpha\gtrsim1$.
Combining the spin temperature with the amount of neutral hydrogen, which is parametrized by the neutral fraction $x_{\rm HI}$, gives the differential brightness temperature
\begin{equation}\label{brightnessT}
T_b=27~ x_{\rm{HI}}\left(\frac{\Omega_bh^2}{0.023}\right)\left(\frac{0.15}{\Omega_mh^2}\frac{1+z}{10}\right)^{1/2}\left(\frac{T_S-T_{\rm CMB}}{T_S}\right)\,\rm{mK}.
\end{equation}

\begin{figure}[htbp]
\begin{center}
\includegraphics[scale=0.35]{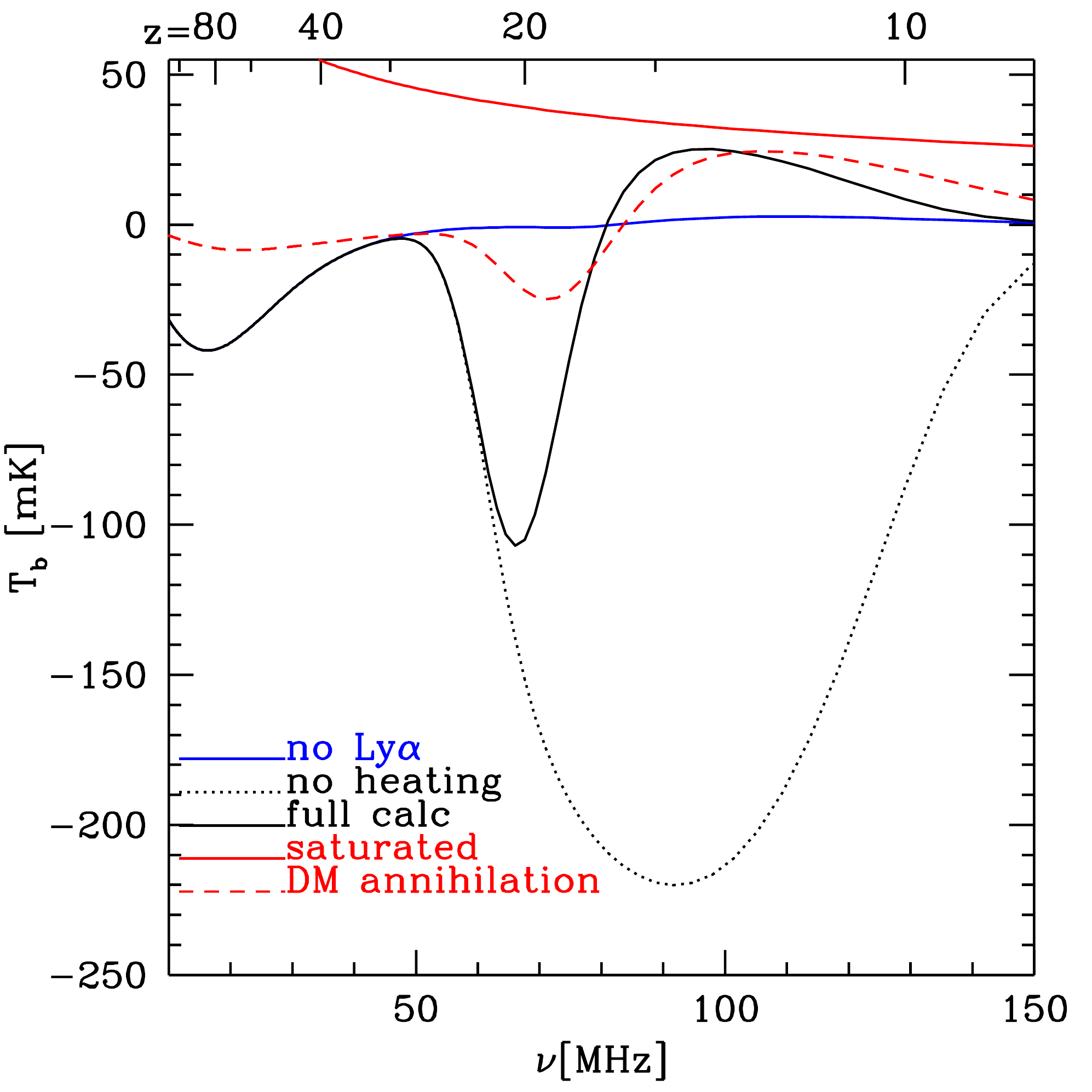}
\includegraphics[scale=0.35]{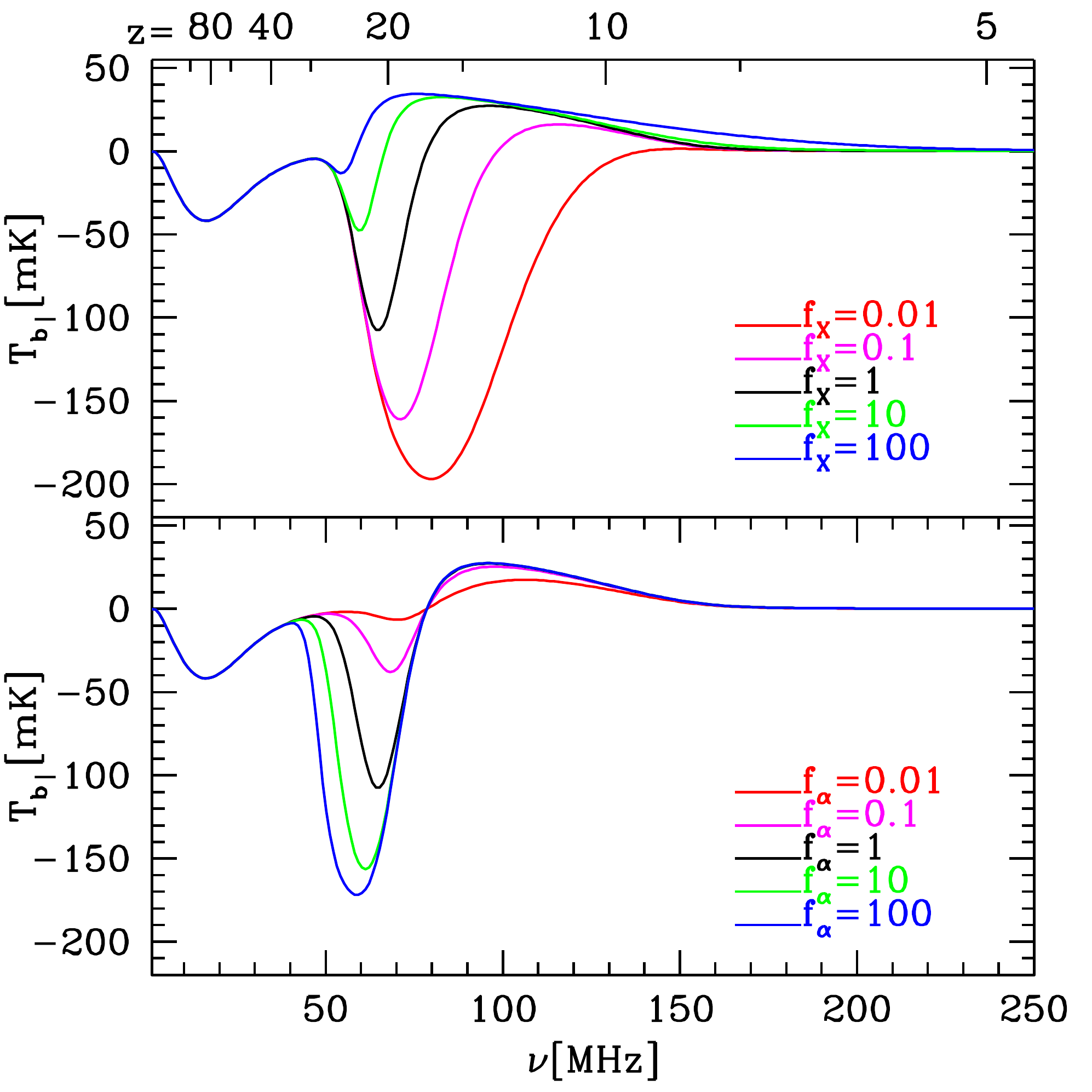}
\label{fig:globalsignal}
\caption{Example histories of the 21~cm all-sky signal. {\em Left panel: }Variation with different physics \cite{2012AdSpR..49..433B} {\em Right panel: }Variation with X-ray heating (top) and \lya flux (bottom) \cite{2010PhRvD..82b3006P}.}
\end{center}
\end{figure}

The evolution of the all-sky 21~cm signal is illustrated in Fig.~\ref{fig:globalsignal} and shows three key regimes - two absorption troughs and one emission feature.

Working chronologically down from high redshift, we first see  a 21~cm absorption signal during the cosmic dark ages. In the absence of exotic new physics, the evolution of this signal is determined by collisional coupling and the effect of adiabatic expansion on the temperature of the gas \cite{Loeb:2004uq}. However, it is possible that exotic physics and dark matter annihilation, for example, can heat the gas in this early phase changing this picture \cite{2012AdSpR..49..433B}.

More generically, collisional coupling becomes ineffective as the gas becomes more diffuse and it is only when the first stars and galaxies form that the 21~cm signal becomes complex. Galaxies produce \lya photons that couple $T_S\approx T_K$ leading to the second absorption feature at redshifts less than about 30. The start of this potentially deep second absorption feature occurs as \lya coupling starts and the end is governed by the eventual heating of the Universe by UV and soft X-rays, which leads to the final emission feature. The right panel of Fig.~\ref{fig:globalsignal} serves to illustrate some of the uncertainty on the all-sky signal with \lya and X-ray emissivity for fixed star formation history and assuming atomic hydrogen cooling. The \lya luminosity is relatively similar for Pop II and Pop III stars and so reasonably well understood. More difficult to predict is the star formation history itself. Recent work has shown that relative velocity between baryons and dark matter can suppress early star formation in low mass halos \cite{Maio:2012kx,2014MNRAS.437L..36F}. The availability of molecular H$_2$ for gas cooling in mini halos is also significant in early star formation.

The production of X-rays and their absorption by the IGM, which is highly dependent on their energy, is still quite poorly understood. Candidate sources include mini-quasars, inverse-Compton scattering from supernova remnants and X-ray binaries in star forming galaxies \cite{2006MNRAS.371..867F}. Current predictions extrapolate based on local source populations, but evolution effects are to be expected \cite{2011A&A...528A.149M}. Moreover, the SED of sources is variable being more or less like a power law at different energies. Hard X-rays (>2~keV) may have mean free paths comparable with the size of the Universe and so fail to heat the IGM. The rising edge of the deep absorption feature provides a way to constrain this physics and has been a topic of considerable recent interest \cite{2008PhRvD..78j3511P, 2013MNRAS.431..621M}.

Ultimately, the IGM is likely to enter a hot, \lya coupled state where $T_S\gg T_{\rm CMB}$, although this is not guaranteed. This saturates the all-sky 21-cm signal removing any dependence on $T_S$, so that $\delta T_B$ is completely determined by the mean density and neutral fraction $x_{HI}$. As ionizing UV photons from star-forming galaxies reionize the Universe the neutral fraction $x_{HI}$ falls to zero. This is the final emission feature of the all-sky 21-cm signal, whose decline indicates the progression of reionization.

Another cosmological signal in the all-sky mean spectrum of the cosmic radio background is recombination lines from the epoch of recombination \cite{Sunyaev:2009fk}.  In contrast to the 21-cm signatures of reionization, the predictions of the recombination spectrum are rather precise \cite{Rubino-Martin:2008fk} assuming standard recombination physics, and hence serve as an excellent test of the standard model.  The recombination lines are expected to appear across SKA bands: in Fig.~2 we show the signature across the SKA-mid frequency range.
\begin{figure}[htbp]
\begin{center}
\includegraphics[scale=0.5]{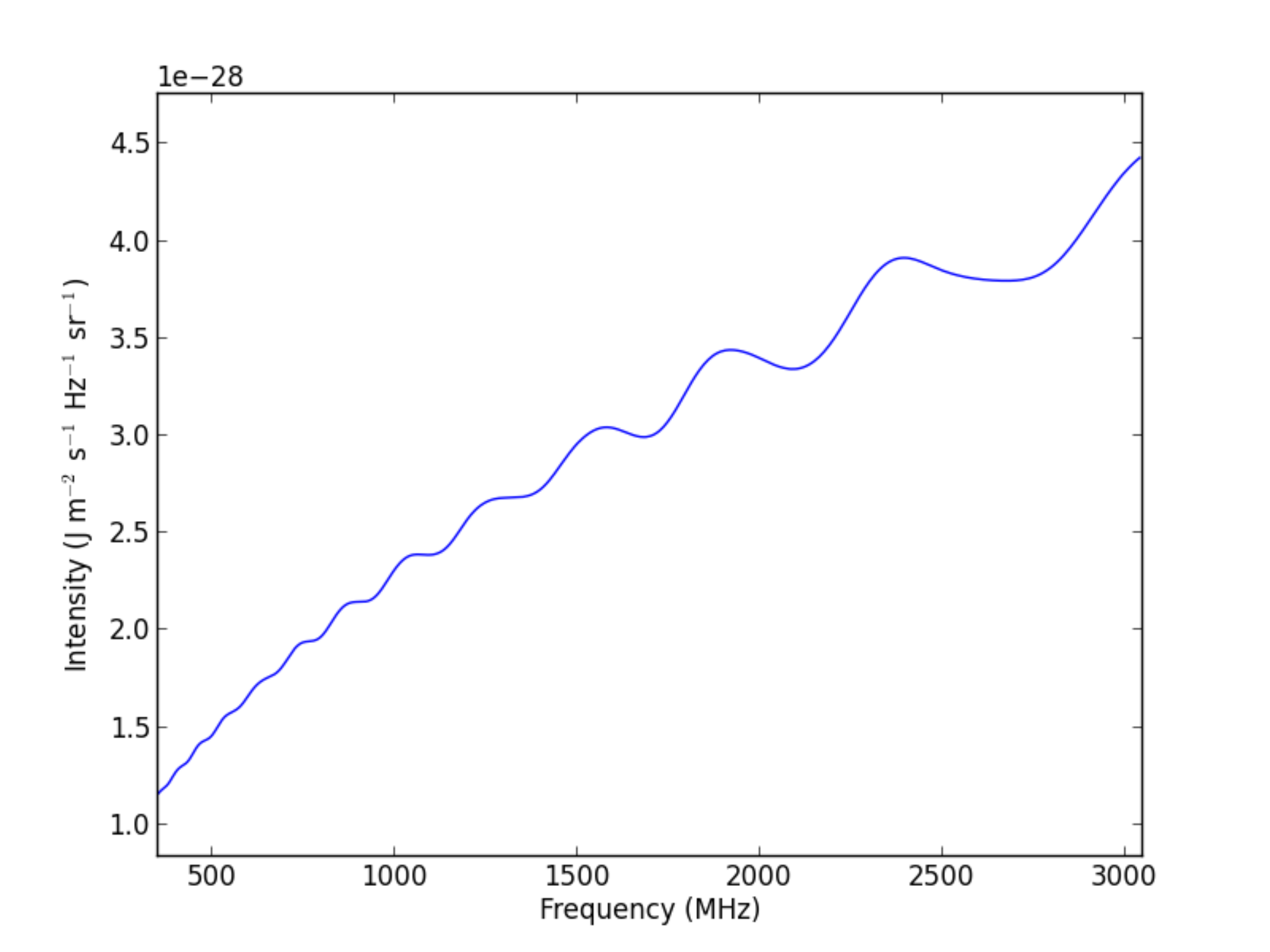}
\caption{Spectrum of recombination lines from the epoch of cosmological recombination \cite{Rubino-Martin:2008fk} across the SKA-mid frequency range.}
\end{center}
\end{figure}
\label{fig:recombinationsignal}

\section{The motivation for all-sky wideband observations with the SKA}

The generic all-sky spectral signature of reionization has an early cosmic time when Lyman-$\alpha$ from the first stars might cause a substantial absorption dip in the cosmic radio background, which is followed by reheating and late time ionization that is expected to manifest as 21-cm emission that progressively diminishes towards lower redshifts.  The spatial distribution of ionizing sources and the density structure of the gas originating in the matter distribution power spectrum results in spatial structure in hydrogen spin temperature and ionization fraction; this manifests as an evolving power spectrum of brightness temperature fluctuations about the evolving mean all-sky absorption or emission signature.  A key science goal of SKA-Low is the interferometer imaging of the cosmic evolution of spatial structure in the ionization state and  21-cm spin-flip excitation temperature.  The detection of the evolving power spectrum of this spatial structure is a diagnostic of the astrophysics in the reionization epoch, the radiation from the first stars and dwarf galaxies, and their distribution statistics.  

In this context it is useful to ask whether the all-sky spectral signature contains any observables that are outside the horizon of the SKA observing in interferometer mode.  The all-sky redshifted 21-cm signature provides, for all redshifts,  the mean departure of the 21-cm brightness temperature from the ambient CMB.  This measurement is the critical zero-spacing measurement of reionization that is vital for establishing the normalization or base level of the interferometer measurement of fluctuation power.  The all-sky signature is a direct measure of the mass-averaged evolution in the state of the gas and establishes clearly whether the interferometer measurements of fluctuating power at any epoch refers to absorption or emission.  The all-sky measurement unambiguously determines the `turning points' in the all-sky spectrum that identifies epochs when the spin state of the gas is driven by Wouthuysen-Field mechanism of coupling to the gas kinetic temperature, when X-ray and UV reheating occur and finally when reionization proceeds.

Today, the relative roles of the UV and X-ray radiations from the earliest stellar populations, dwarf galaxies and galactic nuclei is poorly constrained and as a consequence substantial uncertainty exists in our knowledge of the redshifts at which 21-cm power might be detectable.  Second, the detection of redshifted 21-cm power from these epochs is very dependent on our ability to precisely subtract the confusing power from Galactic and extragalactic foregrounds, which dominate the cosmological signal by orders of magnitude.  For both these reasons, an independent and early detection of the all-sky signature of cosmic evolution in the gas is advantageous in that it aids in selecting optimum bands for SKA imaging of the evolving power spectrum and selecting control bands where the power is expected to be diminished.  SKA measurements of the all-sky signatures are, therefore, to be viewed as an essential precursor to detection and imaging of spatio-temporal 21-cm power and a necessary zero-spacing measurement that provides the base level in which the fluctuation power resides.

In the measurement of the all-sky spectrum of the cosmic radio background with the aim of detecting the faint reionization signatures embedded therein,  the effective collecting area of the sensor of the sky radiation is not of consequence.  What is needed is a spectral radiometer that may be precisely calibrated, and a sensor that is ideally frequency independent so that sky brightness distribution structures do not result in confusing spectral structure in the measurement set.   As expected, there are several ongoing and proposed experiments that aim to make precision measurements of the spectrum of the cosmic radio background: CoRE (Chippendale, PhD thesis, 2009), EDGES \cite{Bowman:2010kx}, LEDA \cite{Greenhill:2012vn}, DARE \cite{Burns:2012ys}, SARAS \cite{Patra:2013zr}, SCI-HI \cite{Voytek:2014ly} and ZEBRA (Mahesh et al. 2014, submitted).  So the question arises as to why we even consider all-sky measurements with SKA elements or stations, when dedicated single-element spectral radiometers that are purpose built for the all-sky measurement might suffice!  

The cosmological signal is embedded in a foreground that is expected to be smooth in the sense that it is a summation over numerous sources within the receptor beam pattern and along the line of sight, each of which have spectra close to a power law but which may have a spread in their indices.  The foreground spectrum would, therefore, be a {\it `complete monotone'} but might require a high order function for precise modeling (Rao et al., in preparation).  The observing band must necessarily include multiple turning points of the predicted reionization signature so that the cosmological signal in the total observing band be indescribable as a complete monotone and hence separable from the foregrounds.  The expected cosmological signatures from reionization are rather wideband signals, and hence separation of the cosmological signal from foregrounds requires observing with bandwidths exceeding an octave. 

The spectral response of any total power spectral radiometer contains the spectrum of the radio background modulated by the spectral gains of the receptor and electronics.   The response includes additive terms from ohmic losses in the receptor of the radiation and its interconnect to amplifier devices, amplifier noise, ground spillover etc.    Depending on the magnitude of the dependence of the receptor beam pattern on frequency, the response also contains spectral structure arising from mode coupling of sky structure into spectral structure.  Over observing bands exceeding an octave, design of receptors with frequency independence of the beam patterns at the precision required to avoid mode coupling is challenging.  However, the SKA in interferometer mode solves simultaneously for sky structure, receptor pattern and hence provides calibration solutions for the mode-coupling; this is the prime motivation for SKA measurement of the all-sky signal.

Recombination lines from the epoch of recombination are expected to appear as an extremely weak ripple in the cosmic radio background.  Their detection is likely best done at cm wavelengths, where the signal-to-noise is expected to be greatest considering the line intensities, background sky noise contribution to measurement errors and receiver technology (Rao et al., in preparation).   Detection of the recombination spectral structure requires observing with about $10^3$ elements for reasonable integration times.  SKA-mid elements operating as spectral radiometers is perhaps the only instrument capable of detecting these features in the foreseeable future; as discussed above their operation as interferometers provides the calibrations for element bandpass and gains as well as solutions for mode coupling.  

\section{Observing all-sky cosmological signals with the SKA}

The dark-ages and reionization signatures are from about 40 to 200 MHz, almost wholly in the band of  SKA-Low Phase 1.  Nominally, the interferometer elements in SKA-Low Phase 1 are phased arrays of 256 dual-polarized log-periodic antennas, referred to as stations. The measurement set for this science is the total power spectra of the interferometer elements of SKA-Low operating as spectral radiometers, which corresponds to the auto-correlation spectra of the station signals.   Recombination lines are predicted to be present across all SKA bands; however, the optimum frequency range for SKA is at the top of the SKA-mid band, and the total power spectra of the dish elements that form the SKA-mid interferometer form the measurement set for that science.

Critical to the science is the calibration of these total-power spectral radiometers.  Calibration consists of (a) determining the instrumental bandpass for the sky signal, (b) modeling the additive spectral contaminations from other sources of system noise: receiver noise, ground spillover etc., and (c) correcting the measurement set for mode-coupling.  The interferometer visibility measurements form the calibration set that generate the all-sky model of the sky brightness distribution while simultaneously solving for a model of the antenna pattern and the interferometer element bandpass and gains.  These provide the bandpass for the sky signal in the station-based total power radiometers, and also provide model of the mode coupling of sky structure into spectral structure.  Certain additive contaminants (receiver noise etc.) require special and separate calibration via engineering design of the interferometer elements. 

Since collecting area is inconsequential for the all-sky signal, and simplicity and frequency invariance in the beam pattern is critical, the measurement set for the reionization signature may be derived using the sky signal from just one log-periodic antenna in each station of SKA-Low.  In this case just that station array element may be outfitted with electronics for calibration of its auto-correlation spectra.  Alternately, a separate precision element may form an outrigger to each station and provide a wide-field auto-correlation signal, which would serve as the measurement set for the EoR science.  It may be noted here that an outrigger that is placed sufficiently isolated from the station elements would escape inter-element coupling effects that are highly chromatic.

\subsection{Interferometric measurement of a global signal using lunar occultation}

Interestingly, the all-sky 21-cm signal can also be measured interferometrically \cite{shaver1999}, thereby circumventing some of the calibration challenges in total-power measurements. Though an interferometer is in general insensitive to a uniform all-sky signal, the presence of an occulting object such as the Moon imposes spatial structure on an otherwise featureless all-sky signal. Interferometers being spatial differencing instruments are sensitive to this structure, and measure the brightness contrast between the occulted background (diffuse Galactic emission plus all-sky $21$-cm signal) and the Moon. The Moon is expected to behave as a $230$~K blackbody (spectrally featureless) at radio frequencies \cite{heiles1963} and provides a suitable constant temperature reference against which the spectral structure in temperature of the uniform background may be measured.  The spectrum observed in synthetic beams of SKA-Low towards the Moon and over the full band of 50-350~MHz provide a difference measurement between the Moon and background, which would contain the reionization spectral structure.  This technique has been demonstrated recently for the first time between $35$ and $75$~MHz using LoFAR \cite{2014arXiv1407.4244V}.

Reflected Earthshine (radio frequency interference or RFI) from the Moon was thought to be a limitation to this technique \cite{mckinley2013}. However, recent work with LOFAR \cite{2014arXiv1407.4244V} has shown that reflected Earthshine images to the center of the lunar disc due to the specular nature of reflection and, therefore, may be independently modeled and subtracted using longer baselines ($>100$ spatial wavelengths) that have the resolution to isolate this contaminant.  Another potential limitation of this technique stems from contamination of lunar flux by Galactic and Extragalactic emission because of mode-coupling due to chromatic sidelobes of the interferometer synthetic beam.   Here again all-sky images of continuum structure that are observed along with solutions for the station gain and bandpass calibrations provide the models for subtracting the mode-coupling.

Apart from the primary cosmic dawn/EoR science goals, the lunar occultation technique is expected to yield extremely interesting early science results from the SKA via the combination of interferometer data on the Moon and total power spectra from absolute calibration of interferometer elements to accurately measure the brightness temperature of the Moon itself. Since the thermal emission from the Moon comes from a depth of $\sim 100$ wavelengths, this technique facilitates an unprecedented  measurement of the depth evolution (if any) of the lunar regolith temperature up to depths of $\sim 1$~km.  

\section{SKA design requirements for the science case}






The SKA1-Low will consist of 1024 stations each containing 256 log-periodic dual-polarized antenna elements evenly distributed in an irregular-random configuration within a station diameter of 35~m.  The stations are distributed in an array configuration that places 50\% within a 600~m radius, 75\% within a 1~km radius, and the remaining 25\% in three spiral arms that give maximum baseline of 100~km.  The array will operate from 50 to 350~MHz.  All of the 256 log-periodic dipoles in each station are usually combined in a beam former and the dual polarization signals from 1024 stations will be transported to a central signal processing building where they will be channelized and cross-correlated to provide full Stokes measurements of the spectral visibilities.  The station beams, when formed from all 256 elements, expose a field-of-view defined by a  circularly symmetric beam of $5^{\circ}$ at the half-power points at 100 MHz. 

We have mentioned in Section~3 above three different and alternate approaches to generating the basic measurement sets for the all-sky reionization signal using SKA-Low. They are (a) to use as basic elements the 35~m diameter stations configured as phased arrays of the 256 dipoles, or (b) to derive the sky signal by enabling just one log periodic element in each station, or (c) to separately add, as an outrigger to each station, an antenna element with precision calibration capability that would provide the autocorrelation spectra.  

In all the three approaches, the observing strategy is to measure the autocorrelations of both polarization signals (X \& Y) from each of the stations---to serve as the measurement set---and also full Stokes correlations (XX, YY, XY \& YX) between all pairs of station signals to serve as the calibration set.  This may be done for only signals from stations in the central core.  An obvious advantage of restricting EoR all-sky measurements to the core stations is that this restricts the baselines used to those that are not severely affected by ionosphere; therefore, element gains, phases and bandpass calibrations may be derived with greater accuracy. The SKA1-Low correlator is required to be configured to generate the total power autocorrelation spectra for the station signals from the central core, as well as the usual interferometer cross correlations.  It may be noted here that the overhead to generate the autocorrelations is relatively small and the autocorrelations may be performed by the resources freed by omitting the correlations between station signals from the outer arms.


Strategy (a) above has the highest collecting area per station and hence maximum sensitivity in baselines; consequently, this option has minimum calibration errors and produces the best global sky model.  It may be noted here that enabling recording of accurate and useful auto-correlation spectra from these station signals is critical to using their sky-scans for providing the short and zero-spacing information, which will be vital for making complete and absolutely-calibrated global sky models.  This is no doubt important for Galactic and Extragalactic science.  From the perspective of all-sky EoR science, making an excellent global sky model with all sky-structure modes represented is critical to subtracting the spectral structures in autocorrelation spectra arising from mode coupling.




Spectral confusion due to any chromatic variation in the beam, which would couple foreground structure to spectrum (the mode-coupling issue), indeed has a solution with the SKA.  In option (a) above, where the entire SKA-Low station is phased to form the signal for the total power detector, the station beams are inevitably frequency dependent.  In this case the relatively superior station sensitivity also has the capacity for creating global sky models with greater accuracy, and as we have discussed above, the stations may also provide measurements of large angular modes to make the global sky model more truly representative of the absolute sky brightness distribution.   A frequency dependent weighting of the element voltages prior to combining to form the station beams could potentially reduce the frequency dependence of the station beams, but not exactly because the elements do not provide continuous coverage of the aperture field. Along with interferometer solutions for the bandpass of the station beams, the models also provide calibrations for the mode coupling.  Option (b) above that deploys single log-periodic dipole elements as the total power detectors would also suffer mode-coupling owing to the frequency dependence in the beam (by definition, the antenna is log-periodic, which implies that it has periodic structure in antenna characteristics in log-frequency space).  In this option the sky model and antenna response modeling that emerges from the interferometer data would provide relatively poorer calibration of the mode coupling.  Option (c) above uses outriggers: these antennas may be designed to be wideband frequency-independent elements with relatively lower mode coupling.   Even if the elements and station beams are frequency independent, it may be mentioned here that the ionosphere could make the effective beam chromatic.

The detection and mitigation of RFI will be far superior in all-sky measurements based on autocorrelation spectra of SKA stations, which simultaneously record interferometer data.  The SKA-Low baselines between the core stations are relatively small in wavelength units and hence RFI is poorly rejected in the interferometer integration times.  Therefore, the visibilities are excellent detectors of RFI, particularly when Stokes V visibilities are formed, and constitute a calibration set for RFI recognition in time-frequency space that may be used to reject RFI-corrupted station signals.  Moreover, interferometer data that do not contain the uniform all-sky signal may also be used to model and subtract RFI from the station total power spectra, which constitute the measurement set.  Real-time RFI rejection and mitigation algorithms that clean the SKA station autocorrelation spectra using visibility data is desirable.
  

Experiments to date with single receptors have shown that the calibration of additive bias in the auto-correlation data is challenging, and many methods have been proposed for their marginalization. The additive contributions to autocorrelation spectra arise from ground spillover, antenna and balun ohmic losses, losses in the analog electronics signal path, and amplifier noise, which has components that propagate along and against the nominal signal flow direction.  Impedance mismatches within the signal path cause reflections and, therefore, these system temperature  components appear in the spectra of measurement sets with frequency structure owing to multi-path propagation to the detector.  The  marginalization of additive bias requires special calibration techniques; some methods are discussed and demonstrated in designs of EDGES and SARAS.  

Using station beams as in option (a) above requires designing all log-periodic dipoles to have additive terms that are smooth functions of frequency that are separable from EoR signatures, which have turning points.   Possible solutions include restricting path lengths for multi-path propagation by, for example, integrating high-gain amplifiers with baluns at the antenna terminals and optically isolating this front-end electronics from the rest of the receiver.  Calibration of the response to amplifier noise requires either that the amplifier noise figure be switchable between high and low states, so that a differential response would provide the response to the amplifier noise.  An alternate method that has been explored is including a bidirectional coupler with the amplifier so that a calibration noise may be injected at the same point in the signal path to measure the spectrometer response to a bidirectional noise that originates at the amplifier.


SARAS demonstrates the value of correlation receivers that yield difference measurements between the sky power spectra and that from an internal reference load.  Switching in the correlation receiver cancels many internal systematics and additive contaminants downstream of the switch.  The correlation receiver provides complex spectra in which the sky power is exclusively in the real part following bandpass calibration, and the additive contaminants from receiver noise appear in real and imaginary parts with quadrature phase offset allowing for modeling and marginalization of their effects.  Switching of the internal load provides a means for separating its contribution.  SARAS represents a purpose built complex calibration scheme that may hence be included in options (b) and (c). 

Bandpass calibration of spectral radiometer elements that form SKA interferometers may be done using spectral visibilities to simultaneously solve for sky structure and element gains.  Switched injection of flat-spectrum noise power via directional couplers is an alternate engineering approach to bandpass calibration---implemented in SARAS---that requires noise sources and couplers in every radiometer element.   An alternate engineering solution for bandpass calibration of the radiometers is pulse calibration, which is the time domain equivalent of flat spectrum noise injection.  By measuring the system response to pulses much narrower than inverse of the bandwidth,  the bandpass calibration is derived and also different modes in the observed spectra may be tagged to reflections from different impedance discontinuities in the signal path.  Yet another method for bandpass calibration is the injection of a swept frequency tone into the system path.

In addition to bandpass calibration and the marginalization of additives, precision polarimetric calibration to estimate the unwanted confusing leakages from the polarized structure in the sky into spectral modes is necessary.  Linearly polarized sky emission will appear with spectral structure in total power spectra of linearly polarized antenna elements if Faraday rotation changes the plane of polarization arriving at the antenna over frequency.  Improper instrumental calibration
along with Faraday Rotation that causes the polarized signal to have frequency structure might confuse the all-sky EoR signature. In this context, once again we see the advantage of using the SKA for all-sky EoR signal detection because of the ability of an interferometer array to also yield solutions for precision polarimetric calibration for the interferometer elements, which in this case constitute the spectral radiometers that measure the all-sky 21-cm signature.  Specifically, Faraday rotation-measure (RM) synthesis using interferometer visibilities offers a more complete and hence accurate polarimetric calibration, associating each pixel in the sky with Stokes I along with (Q, U, RM) sets and their spectral indices. 
The Stokes V data are also useful in setting the thermal noise level apart from providing an indicator for radio frequency interference.
The off-axis polarization effects are smaller in strategy (a), which uses a narrow station beam of $5^{\circ}$, compared to strategies (b) and (c), which use single receptors with broad beams as basic elements. 

All of the above discussions on the requirements for interferometer based calibration of element bandpass, and solving for the mode coupling of sky structure to spectral structure from continuum radio sources as well as linearly polarized emission along with Faraday rotation, are also relevant to SKA-mid for detection of recombination lines.  The SKA-mid antennas are dishes that form the interferometer elements: their autocorrelation spectra form the measurement set for the detection of recombination lines are hence the additives in these signals will require calibration.  Dish-type antennas are electrically large structures and hence would have long lengths in the signal path arising from reflections off structural elements.  For this reason, for the all-sky science it is desirable for the dishes to be off-axis reflector antennas.

\section{The path to SKA-Low detection of all-sky 21-cm from reionization}

It is often the case that effective use of a complex experimental system for a key science goal  for which it is custom built requires a deep understanding of performance and systematics, which requires experience with the instrument and development of appropriate calibration methods in the commissioning phase.  New telescopes that push frontiers of technology come with surprises beyond the imagination of its designers and builders.  

The sensitivity of a spectral radiometer to the all-sky EoR signature is independent of collecting area and depends on the system temperature and integration time.   Pilot observations, which could lead to useful science, could begin early in the commissioning phase with 128 stations once interferometer imaging to solve for the global sky model and system calibration is enabled. The expected system temperature at 110 MHz is about 800~K and this leads to the nominal requirement of about 1~hr observing time with 128 station beams to achieve 1~mK sensitivity with a 1~MHz bandwidth.   The actual time required to achieve this sensitivity would be a factor of a few larger because of various switching schemes that are inevitably introduced for calibration.  For example, EDGES has a three-way switching scheme while SARAS cycles through six states of the receiver to obtain measurement and calibration data products.  The above bandwidth and sensitivity are adequate to characterize the turning points in the spin temperature evolution curve (see Fig.~1).   The advantage as the buildout progresses towards completion of SKA Phase 1 is not in thermal-noise related sensitivity to all-sky EoR signal, but in increasing accuracy of the calibration of the individual radiometer elements.

It may be noted here that the foregrounds are of substantially greater brightness towards longer wavelengths since the cosmic radio background has a steep $T_{b} \sim \nu^{-2.5}$ dependence on frequency; therefore, successful imaging of the spatial structure in 21-cm spin flip is increasingly challenging towards earlier times in the evolution of the gas.  The substantially greater calibration accuracy and confusion subtraction required for elucidating early-time evolution through interferometer imaging encourages designing for detection first of the wideband all-sky signature, which may be done in SKA Phase 1. All-sky detection of the absorption trough will serve as a motivation and prelude to later imaging of the spatial structure. 

The experience with SKA Phase 1 would be critical to the design of the elements of SKA Phase 2: in the design of the antenna elements and calibration schemes for additives.

\section{The path to SKA detection of the recombination era}

The all-sky recombination spectral signature requires $10^3$ spectral radiometers operating in parallel, and whose calibrated spectra are co-added, for detection in reasonable time.  As demonstrated in Rao et al. (in preparation), arrays of radiometers with about octave bandwidths within Band~4 of the SKA-mid frequency range would require about 25,000~antenna-days for a $1\sigma$ detection.  SKA1-mid array of 190 antennas, with dual polarization, would require 66 days for such a detection.   A genuine attempt at a detection of the recombination spectrum is really reserved for the full SKA, after completion of SKA2-mid in Phase 2.   Long duration imaging campaigns, perhaps as a deep all-sky survey, would yield the data set that might serve for detection of the recombination line signature with SKA in Phase 1 and in Phase 2.  

\bibliographystyle{unsrt}
\bibliography{allrefs}{}









\end{document}